\documentstyle[preprint,aps]{revtex}

\newcommand{\be}{\begin{equation}}
\newcommand{\ee}{\end{equation}}

\newcommand{\ba}{\begin{array}}
\newcommand{\ea}{\end{array}}

\newcommand{\bea}{\begin{eqnarray}}
\newcommand{\eea}{\end{eqnarray}}

\begin{document}

\title{ Quantum Dynamics of Topological Singularities: \\ 
     Feynman's Influence Functional Approach  }

\author{ P. Ao }

\address{ Department of Theoretical Physics, Ume\aa{\ }
        University, 901 87 Ume\aa, Sweden \\E-mail: ao@tp.umu.se }

%\author{Elsie Tan,  Jessie Tan and R. Sankaran}

%\address{World Scienfitic Publishing Co Ltd, 
%57 Shelton Street, Covent Garden, London WC2H 9HE, England\\
%E-mail: wspc@wspc.ox.uk}  


\maketitle

\abstract{
 Starting from the microscopic theory of Bardeen-Cooper-Schrieffer (BCS) 
 for the fermionic superfluids, we show that the vortex dynamics can
 be followed naturally by extending Feynman's influence functional approach
 to incorporate the transverse force.
 There is a striking mutual independence 
 of the transverse and longitudinal influences:
 The former has the topological origin and is insensitive to details, while
 the latter corresponds to the well-known damping kernel depending on
 details.   
}

\section{ Introduction }

Vortices in superconductors and He3 superfluids are topological 
excitations. They determine the global properties such as
the stability of the supercurrent carrying states, 
and have been under intensive theoretical and experimental 
studies since earlier sixties\cite{brandt}.
The detailed microscopic BCS type theory
for fermionic superfluids is successful and well defined. 
The derivation of vortex dynamics is, however,
non-trivial and less certain.  
Here we present a path integral derivation of vortex 
dynamics following the line of
geometric methods\cite{ao1,thouless}.
The advantage of the present method is that in the effective vortex Lagrangian
the separation of geometric and dissipative contributions 
is natural. The important results here are  
that the transverse force agrees with the one obtained by 
the Berry phase method\cite{ao1} and by the total
force-force correlation function method\cite{thouless}, 
and is insensitive to details. 
The friction is determined by the
spectral function of the Hamiltonian, sensitive to details. 
The large transverse force has been found recently by 
a direct measurement\cite{zhu}, in consistent with the prediction in
Ref's.[2,3].

\section{ Influence Functional Approach  }

To find the effective vortex action,
we begin with the standard BCS Lagrangian for s-wave pairing
in the imaginary time representation. Here 
the unwanted fermionic degrees of freedoms will be integrated.
We will only consider neutral superfluids here,
but the coupling to electromagnetic fields does no affect our main results. 
More detailed analysis will be published elsewhere.\cite{az}
The model BCS  Lagrangian is
\begin{eqnarray}
   L_{BCS} & = &  \sum_{\sigma}\psi^{\dag}_{\sigma}( x,\tau) 
      \left(  \hbar\partial_{\tau}- \mu_F 
    - \frac{\hbar^{2}}{2m} \nabla^{2} + V(x)  \right) \psi_{\sigma}(x,\tau)
       \nonumber \\
      & & - g\psi^{\dag}_{\uparrow}(x,\tau) \psi^{\dag}_{\downarrow}(x,\tau)
          \psi_{\downarrow}(x,\tau) \psi_{\uparrow}(x,\tau )  \; , 
\end{eqnarray}
where $\psi_{\sigma}$ describes electrons with spin $\sigma=(
\uparrow, \downarrow )$, $\mu_F$ the chemical potential determined by the 
electron number density, $V(x) $ the impurity potential, and $x =(x,y,z)$.
A vortex at $x_v$ has been implicitly assumed. 
The partition function is 
\be
   Z =  \int {\cal D}\{x_v, \psi^{\dag}, \psi \} \times             
           \exp\left\{ - \frac{1}{\hbar} \int_0^{\hbar\beta } 
             d\tau \int d^3x L_{BCS} \right\}   \; ,      
\ee
with $\beta = 1/k_B T $ , and $d^3x = dxdydz$.
Next, we first 
perform the Bogoliubov transformation by introducing the auxiliary 
fields $\{ \Delta^{\ast}, \Delta \}$,  integrate over the electron
fields $\psi_{\sigma}^{\dag}$ and $ \psi_{\sigma}$, then over 
the auxiliary(pair) fields under the meanfield approximation. 
The partition function for the vortex is  
\be   
   Z = \int {\cal D}\{ x_v \}
       \exp \left\{ - \frac{ S_{eff} }{ \hbar }  \right\} \; ,
\ee 
with the effective vortex action
\be
   \frac{S_{eff} }{\hbar} = - Tr \ln G^{-1} -
     \frac{1}{\hbar g}\int_0^{\hbar\beta} d\tau \int d^{3}x|\Delta|^{2} \; ,
\ee
where $Tr$ includes internal and  space-time indices,
and the Nambu-Gor'kov Green's function $G$ defined by 
\be
     ( \hbar \partial_\tau + {\cal H})
      G(x,\tau; x',\tau') = \delta(\tau-\tau') \delta^3(x-x') , 
\ee
together with the BCS gap equation, or the self-consistent equation,
\be
     \Delta(x,\tau) = -  g \;   
   < \psi_{\downarrow}(x,\tau) \psi_{\uparrow}(x,\tau) > \; .
\ee
%A special attention should be paid to the equal time limit of the 
%NG Green's function.\cite{schrieffer}
Here the Hamiltonian is defined as
\be
   {\cal H}( \Delta, \Delta^{\ast} ) = \left( \begin{array}{cc} 
                     H & \Delta  \\
                    \Delta^{\ast} & - H^{\ast} 
                   \end{array} \right) \; ,
\ee
with $H =  - (\hbar^{2}/2m ) \nabla^{2} -  \mu_F + V(x)$.

We assume that the vortex is confined to move in a small regime around a point 
at $x_0$, 
which allows a small parameter expansion in terms of the difference 
between the vortex position $x_v$ and $x_0$. 
We look for the long time behavior of vortex dynamics under this small
parameter expansion. 
For the meanfield value of the order parameter, this expansion is
\be
   \Delta(x,\tau, x_v) =  \left( 1 
         + \delta x_v(\tau) \cdot \nabla_0 
       + \frac{1}{2}  ( \delta x_v(\tau) \cdot \nabla_0 )^2 \right) 
         \Delta_0(x,x_0) \; .
\ee
Here $\delta x_v = x_v - x_0$.
In Eq.(8) we have used the fact that when $x_v = x_0$ 
the vortex is static.
The effective vortex action to the same order is, 
after dropping a constant term and
assuming the rotational symmetry under the impurity average,
a straightforward calculation leads to the following effective vortex action 
\bea
    S_{eff} & = & \left. \frac{1}{2}\int_0^{\hbar\beta} d\tau 
     \right[ K \; |\delta x_v(\tau) |^2 
               +  \int_0^{\tau } d\tau' 
      F_{\parallel} ( \tau-\tau')|\delta x_v(\tau)- \delta x_v(\tau') |^2 
       \nonumber \\
           & &   + \left.  \int_0^{\hbar\beta} d\tau' 
            F_{\perp} ( \tau-\tau') 
           (\delta x_v(\tau) \times \delta x_v(\tau') )\cdot \hat{z} 
                  \right]  \; ,
\eea
with the spring constant in the effective potential,
\be
    K =  \frac{1}{g} \int d^3x |\nabla_0 \Delta_0^{\ast}(x,x_0)|^2
        - \int^{\infty}_{0} d\omega \frac{ J(\omega) }{\omega} \; ,
\ee
the damping kernel,
\be
    F_{\parallel}(\tau) = \frac{1}{\pi} \int^{\infty}_{0} d\omega J(\omega)
    \frac{\cosh\left[\omega\left(\frac{\hbar\beta }{2}-|\tau|\right)\right] }  
         {\sinh\left[\omega \frac{\hbar\beta }{2} \right]          }  \; ,
\ee
and the transverse kernel, in the long time limit,  
in terms of the virtual transitions,
\bea
   F_{\perp}(\tau) & = & - \partial_{\tau } \delta(\tau ) 
                       \sum_{k,k'} \int d^3x \int d^3x' \;
        \hbar ( f_k - f_{k'} ) \nonumber \\
    & &   \frac{1}{2} \hat{z}\cdot
       \left(  \Psi_k^{\dag}(x') \nabla_0\Psi_{k'}(x')
        \times \nabla_0\Psi_{k'}^{\dag}(x)  \Psi_k(x) \right) \; ,
\eea
or in terms of the contribution from each state,
\bea
   F_{\perp}(\tau) & = & - \partial_{\tau } \delta(\tau) \sum_{k}
                                    \int d^3x \; \hbar \hat{z}\cdot 
  \left( f_k \nabla_0u_k^{\ast}(x)\times \nabla_0 u_k(x) \right. \nonumber \\
    & &   
   \left.  - (1-f_k) \nabla_0v_k^{\ast}(x)\times \nabla_0 v_k(x) \right)  \; .
\eea
with the Fermi distribution function $f_k = 1/(1 + e^{\beta E_k} )$, and the 
spectral function
\bea
   J(\omega) & = & \frac{\pi }{2} \sum_{k,k'} 
     \delta(\hbar \omega - | E_k - E_{k'} |)
     |f_k - f_{k'}|\times \nonumber \\
   & & 
     \left|\int d^3x \Psi_k^{\dag}(x)\nabla_0{\cal H}_0 \Psi_{k'}(x)
         \right|^2 \; .
\eea

In Eqs.(12-14), the wavefunctions 
  $\left\{ \Psi_k(x) = \left( \begin{array}{c} u_k(x) \\ v_k(x) \end{array} 
       \right) \right\}$   
are  the eigenfunctions of the Hamiltonian 
${\cal H}_0 = {\cal H}(\Delta_0, \Delta^{*}_0) $, 
determined by the stationary 
Schr\"{o}dinger equation, the Bogoliubov-de Gennes equation, 
\be
   {\cal H}_0 \Psi_k(x) = E_k \Psi_k(x) \; .
\ee
 
%Comparing Eq.( ) with quantum dissipative 
%dynamics in Ref.[6],  
%we have the general expression for vortex friction.
%We discuss more below on the new feature of treating the transverse
%force from the influence functional.
%%%%%$K$, $ F_{\parallel}$, and $ F_{\perp}$ one by one, and show
%that Eq.(  ) contains both the dissipative effect
%and the transverse force, or the Berry phase.

\section{ Discussions}

The influence functional approach has been successfully 
implemented in recent studies of quantum dissipative dynamics\cite{leggett},
as also manifested in the present conference proceedings.
However, only the longitudinal response has been considered in those
systems.
The above microscopic derivation of vortex dynamics demonstrates 
that the transverse force can be obtained from this approach, too. 
Therefore we have extended the influence functional approach.

Another interesting feature of the transverse response is its 
insensitivity to details, which 
is most transparent from Eq.(13):
In the one-body density matrix both the electron number 
density and the phase $\theta(x - x_v)$  
are all insensitive to details, if the localization effect caused by
impurities is negligible.
Here the phase $\theta$ is defined through the order parameter 
$\Delta(x,\tau, x_v) \rightarrow  |\Delta| e^{ i q \theta(x-x_v) } $, with 
$q = \pm 1 $ describing the vorticity along the $\hat{z}$ direction 
and $\theta(x) = \arctan(y/x)$.  

%\acknowledgments
{ This work was supported by Swedish NFR. }

\end{document}